\documentclass[aps,pra,twocolumn,showpacs,letterpaper,superscriptaddress]{revtex4-1}

\usepackage[colorlinks=true, citecolor=blue, linkcolor=blue, urlcolor=blue]{hyperref}

\usepackage{graphicx,dcolumn,longtable,epsfig}
\usepackage[usenames]{color}
\usepackage{amssymb}
\usepackage{amsmath}
\usepackage{epstopdf}
\usepackage{bm}
\usepackage{footnote}
\usepackage{float}
\usepackage{subfigure}
\usepackage{color}
\usepackage{ulem}

\input epsf.tex

\def\bea{\begin{eqnarray}}
\def\eea{\end{eqnarray}}
\def\be{\begin{equation}}
\def\ee{\end{equation}}

\begin{document}
\title{The effect of disorder on the phase diagrams of hard-core lattice bosons with cavity-mediated long-range and nearest-neighbor interactions}
\author{ Chao Zhang}
\affiliation{Theoretical Physics, Saarland University, 66123, Saarbr$\ddot{u}$cken, Germany}
\author{Heiko Rieger}
\affiliation{Theoretical Physics, Saarland University, 66123, Saarbr$\ddot{u}$cken, Germany}

\begin{abstract}
We use quantum Monte Carlo simulations with the worm algorithm to study the phase diagram of a two-dimensional Bose-Hubbard model with cavity-mediated long-range interactions and uncorrelated disorder in the hard-core limit. Our study shows the system is in a supersolid phase at weak disorder and a disordered solid phase at stronger disorder. Due to long-range interactions, a large region of metastable states exists in both clean and disordered systems. By comparing the phase diagrams for both clean and disordered systems, we find that disorder suppresses metastable states and superfluidity. We compare these results with the phase diagram of the extended Bose-Hubbard model with nearest-neighbor interactions. Here, the supersolid phase does not exist even at weak disorder. We identify two kinds of glassy phases: a Bose glass phase and a disordered solid phase. The glassy phases intervene between the density-wave and superfluid phases as the Griffiths phase of the Bose-Hubbard model. The disordered solid phase intervenes between the density-wave and Bose glass phases since both have a finite structure factor. 

\end{abstract}

\maketitle

\section{Introduction}

The study of adding disorder to the interacting many-body bosonic systems attracts enormous attention both experimentally  and theoretically~\cite{Inguscio2010, DeMarco2009PRL, DeMarco2010Nature, Schneble2011, DErrico:2014gc, Rapsch:1999bw, Ceperley1991, Soyler:2011ik, Ceperley2011, Zhang:2015it, Zhang:2017ei, deAbreu:2018gc, Niederle:2013jy, Zhang:2018ew}. 
Experimentally, ultracold atoms in optical lattices are a promising tool to study quantum phases and quantum phase transitions in strongly correlated quantum many-body systems. It provides an unique possibility of engineering matter with an unprecedented level of control over the parameters entering the Hamiltonian. On one hand, short-range interactions can be realized using Feshbach resonances, while long-range interactions have been studied using ultracold gases of particles with large magnetic or electronic dipole moments~\cite{Booth:2015fp, DePaz:2013ff, Lu:2011hl},  polar molecules~\cite{Yan:2013fn, Hazzard:2014bx}, atoms in Rydberg states~\cite{Saffman:2010ky, Lw:2012ct, Gunter:2013fv}, or cavity-mediated interactions~\cite{Baumann:2010js, Landig:2016il}. On the other hand, disordered potential can be introduced artificially into the ultracold atomic gases in optical lattices. Speckle patterns is the most commonly used to produce the random potential~\cite{CreatDisorder1, DeMarco2009PRL, AAspect2012}. %The disordered potential can be generated using the optical speckle fields~\cite{CreatDisorder1, DeMarco2009PRL, AAspect2012} as well as bichromatic lattices~\cite{Fallani:2007ir}. 
Bichromatic lattices~\cite{Fallani:2007ir}, introduction of localized atomic impurities~\cite{Schneble2011}, and holographic techniques which produce point-like disorder~\cite{Morong:2015gha} are also used to engineer disorder experimentally.

The effect of disorder on the phases and phase transitions of quantum many-body systems triggered many theoretical studies~\cite{Pollet:2013cp, Vojta:2018go, Vojta:2013hb}. The disordered Bose-Hubbard model (DBHM) allows one to study the interplay between disorder and interactions of ultracold bosons in optical lattices. %In the absence of disorder, the BHM features two phases: superfluid (SF) phase and Mott insulator (MI) phase. 
In the phase diagram of the DBHM, the gapless Bose glass (BG) phase, characterized by a finite compressibility and absence of an off-diagonal long-range order, always intervenes as a Griffiths phase between the superfluid (SF) and Mott insulator (MI) phases~\cite{Pollet2009BG, Gurarie:2009it}. The original DBHM focuses on short-range on-site interactions~\cite{Fisher1}. Recently long-range interactions starts attracting the focus of theoretical research. In the absence of disorder, with long-range interactions, the BHM exhibits a richer phase diagram with additional density wave (DW) and supersolid (SS) phases~\cite{Habibian:2013kw, Habibian:2013eh, Niederle:2016fi, Flottat:2017gn}. The ground state phase diagram of the extended BHM with cavity-mediated long-range interactions has been investigated extensively with the help of mean-field theory~\cite{Keller_2017, Li:2013bn, Niederle:2016fi, Dogra:2016hy, Chen:2016kv}, Gutzwiller ansatz~\cite{Sundar:2016ie, Flottat:2017gn}, quantum Monte Carlo~\cite{Habibian:2013eh, Dogra:2016hy, Flottat:2017gn}, and Variational Monte-Carlo~\cite{Bogner:2019ij} methods in 1D, 2D, and 3D. 
The addition of disorder to the BHM with long-range interactions leads to additional phases. In our recent study, we found that in the DBMH, long-range interactions enhance the supersolid phase~\cite{Zhang:2019uy}. However, a study of the DBHM with cavity-mediated long-range interactions in the hard-core limit is still lacking. In the hard-core limit and without disorder, equilibrium phases of lattice bosons with cavity-mediated long-range interactions were investigated in~\cite{Igloi:2018ig, Blass:2018iw} in 1D where the result shows that the checkerboard supersolid does not exist. While in 2D with nearest-neighbor interactions~\cite{Batrouni:2000cd}, the result shows that the checkerboard supersolid is unstable. However, in the presence of disorder and with hard-core limit, whether the supersolid phase exists or not is still unknown.

% The DBHM with the dipolar interaction have been studied ~\cite{Zhang:2017ei}, where at half-filling, the system has SF, BG, and density wave (DW) phase. 

%Without the on-site repulsive interaction, whether the SS phase exists or not is still unknown. 

In this paper, we use quantum Monte Carlo simulations with the worm algorithm~\cite{Prokofev:1998gz} to study the phase diagram of the two-dimensional disordered Bose-Hubbard model with cavity-mediated long-range and nearest-neighbor interactions in the hard-core limit. The paper is organized as follows: in section~\ref{sec:sec2}, we introduce the Hamiltonian of the system of hard-core bosons with cavity-mediated long-range and nearest-neighbor interactions in the presence of disorder. In section~\ref{sec4.1}, we present the phase diagrams of hard-core bosons in the two-dimensional lattice with cavity-mediated long-range interactions for both clean and disordered systems. We also study the phase diagram of hard-core bosons in the two-dimensional lattice with nearest-neighbor interactions for both clean and disordered systems in section~\ref{sec4.2}. Section~\ref{sec:sec5} concludes this paper. 

\section{Hamiltonian}
\label{sec:sec2}
We consider bosons trapped in an optical lattice with both short-range on-site and cavity-mediated long-range interactions in the presence of disorder in the hard-core limit. The bosons are trapped in a two-dimensional (2D) square lattice with linear size L and periodic boundary conditions. 
The hard-core limit corresponds to large on-site interactions where the occupation of two bosons on the same lattice site is suppressed. %At integer filling $\langle n \rangle =1$ and in the absence of disorder, the system is either in an SF state at lower dipolar interaction strength or in a checkerboard (CB) solid phase at larger dipolar interaction strength~\cite{CapogrossoSansone:2010em}. 
The system is described by the Hamiltonian~\cite{Habibian:2013eh, Habibian:2013kw, Niederle:2016fi}:
\begin{align}
\nonumber H& =-t\sum_{\langle i, j\rangle }(a_i^\dagger a_j+a_i a_j^{\dagger}) \\
&-\frac{U_l}{L^2} \Big{(}\sum_{i \in e} n_i - \sum_{j \in o} n_j \Big{)} ^2+ \sum_i (\varepsilon_{i}-\mu)n_i \;\; .
\label{Eq1}
\end{align}

Here the first term is the kinetic energy characterized by the hopping amplitude $t$. $\langle \cdots \rangle$ denotes nearest neighboring sites, %on an underlying square lattice of linear size $L$ with periodic boundary conditions, 
$a_i^\dagger$ ($a_i$) are the bosonic creation (annihilation) operators satisfying the usual bosonic commutation relations and $n_i=a_i^{\dagger}a_i$ is the particle number operator. The second term is the cavity-mediated long-range interaction with interaction strength $U_l$, the summations $i \in e$ and $j \in o$ denote summing over even and odd lattice sites, respectively~\cite{Niederle:2016fi}. The third term is the chemical potential term with chemical potential $\mu$ shifted by the on-site random potential $\varepsilon_{i}$, where $\varepsilon_{i}$ is uniformly distributed within the range $[-\Delta, \Delta]$. $\Delta$ is the disorder strength. The hard-core condition $a_i^{\dagger 2}=0$ implies that sites with more than one atom are energetically suppressed due to a large on-site interaction energy penalty. In this limit the usual on-site interaction term of the Bose-Hubbard model does not play any role. The maximum atom per site is one. The unit of energy and length are set to be the hopping amplitude $t$. For each $\mu/U_l$, $t/U_l$, and $\Delta/t$, we average over 200-400 realizations of disorder.

%From the mean-field level~\cite{Dogra:2016hy}, the cavity-mediated long-range interaction and the nearest-neighbor interaction are equivalent to each other up to a renormalization of the chemical potential and on-site interaction. In order to check this, 

We also consider the extended Bose-Hubbard model with nearest-neighbor interactions in the presence of disorder in the hard-core limit. The Hamiltonian is written as: 
\begin{align}
\nonumber H& =-t\sum_{\langle i, j\rangle }(a_i^\dagger a_j+a_i a_j^{\dagger})\\
&+U_{nn} \sum_{\langle i, j\rangle} n_i n_j+ \sum_i (\varepsilon_{i}-\mu)n_i \;\; .
\label{Eq2}
\end{align}
Here, the first term is the kinetic energy with hopping amplitude $t$. The second term is the repulsive interaction with interaction strength $U_{nn}$ between bosons on nearest neighboring sites. The third term is the random potential term coupled with the chemical potential term. For each  $\mu/U_{nn}$, $t/U_{nn}$, and $\Delta/t$, we average over 1000-3000 realizations of disorder. 

%Without cavity-mediated long-range interaction $U_l=0$ and nearest-neighbor interaction $U_{nn}=0$, the models~\ref{Eq1} and~\ref{Eq2} reduce to the well known disordered Bose-Hubbard model. The ground state phase diagram has been studied in Ref~\cite{Soyler:2011ik}. At filling factor $\langle n \rangle =1$, there exists three phases: the SF phase, the MI phase, and the BG phase. %Including the cavity-mediated long-range interaction, $U_l$, leads to new phases. The density-wave phase, and the supersolid phase. 
%On the other hand, in the absence of disordered potential, phase diagrams of the system with cavity-mediated long-range interaction and nearest-neighbor interaction at various filling factors have been studied by Variational Monte Carlo method in Ref~\cite{Bogner:2019ij}. There are various phases exists: the SF phase, the MI phase, the DW phase, and the SS phase. When disorder presents in the above system, the glassy phase, characterized by finite compressibility and absence of off-diagonal long-range order, emergences. 

%In the following, we perform large-scale quantum Monte Carlo simulations with the worm algorithm~\cite{Prokofev:1998gz} to study equilibrium phases of models~\ref{Eq1} and \ref{Eq2}.

%Table~\ref{Table1} summarizes the order parameters for the various phases we discussed above. 

\section{Ground state phase diagrams}
In this section, we present the ground state phase diagram of the extended BHM with cavity-mediated long-range interactions $U_l$ (model~\ref{Eq1}) in section~\ref{sec4.1} and nearest-neighbor interactions $U_{nn}$ (model~\ref{Eq2}) in section~\ref{sec4.2} in the hard-core limit for both clean and disordered systems, respectively. To obtain the phase diagram for cavity-mediated long-range interactions (Fig.~\ref{FIG4}) and nearest-neighbor interactions (Fig.~\ref{FIG1}), we measure the superfluid stiffness, compressibility, and structure factor to separate those quantum phases. Different phases can be distinguished by the different combinations of those order parameters. Table~\ref{Table1} shows quantum phases and corresponding parameters associated to the phase we find in phase diagrams Fig.~\ref{FIG4} and Fig.~\ref{FIG1}.

\begin{table}[h]
\includegraphics[trim=0.5cm 6cm 7cm 3cm, clip=true, width=0.48\textwidth]{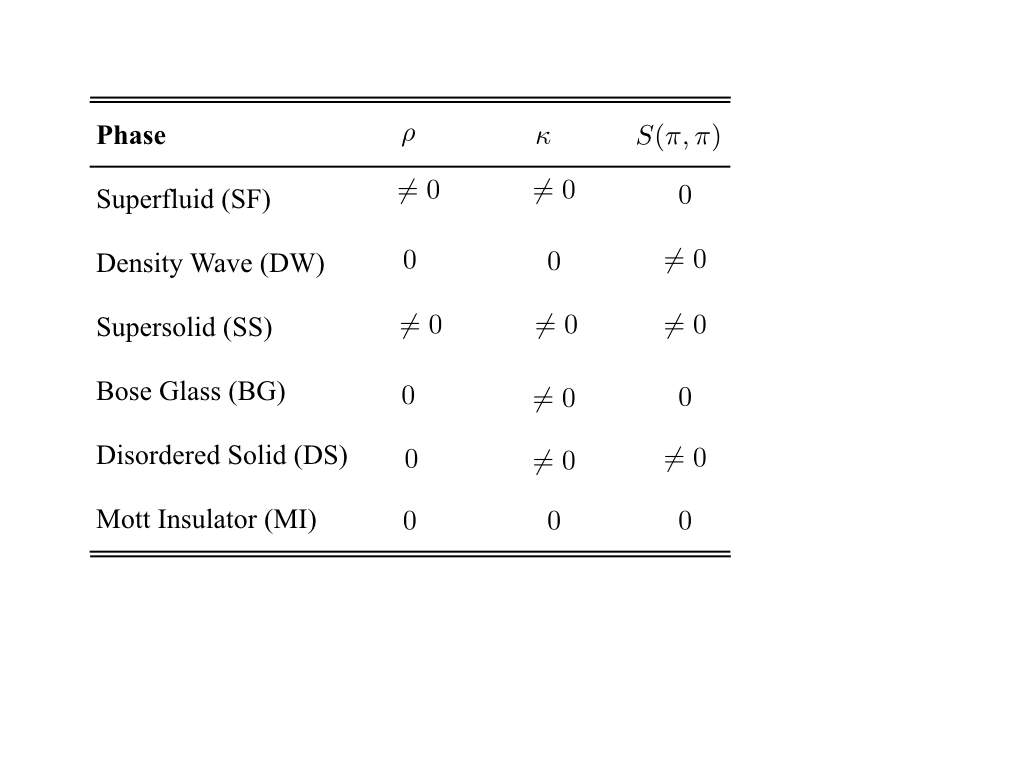}
\caption{Quantum phases and the corresponding parameters: superfluid stiffness $\rho$, structure factor $S(\pi, \pi)$, and compressibility $\kappa$.}
\label{Table1}
\end{table}

The superfluid (SF) phase is characterized by a finite superfluid stiffness, which is easily accessible in the QMC simulations by calculating the winding number $\rho=\langle \mathbf{W}^2 \rangle /dL^{d-2}\beta$~\cite{Winding}. Here, $\mathbf{W}$ is the winding number. $d=2$ is the dimension of the system, $L$ is the linear size of the system, and $\beta$ is the inverse temperature. The density wave (DW) phase has a finite structure factor which is defined as $S(\mathbf{k})=\sum_{\mathbf{r},\mathbf{r'}} \exp{[i \mathbf{k} (\mathbf{r}-\mathbf{r'})]\langle n_{\mathbf{r}}n_{\mathbf{r'}}\rangle}/N$. Here, $\mathbf{k}$ is the reciprocal lattice vector and $\mathbf{k}=(\pi, \pi)$. The supersolid (SS) phase possesses both the diagonal long-range order and off-diagonal long-range order, and it is characterized by finite $\rho$ and $S(\pi, \pi)$. Both the disordered solid (DS) phase and the Bose glass (BG) phase are characterized by a finite compressibility, the difference between them is that DS phase has a finite structure factor. The compressibility measures the density fluctuations and it is defined as: $\kappa=\beta (\langle n^2\rangle -\langle n \rangle ^2 )$. The Mott insulator (MI) phase has zero superfluid stiffness, zero compressibility, and zero structure factor.

\subsection{Cavity-mediated long-range interaction}

\label{sec4.1}

\begin{figure*}[th]
\includegraphics[trim=4cm 0cm 0cm 0cm, clip=true, width=0.75\textwidth]{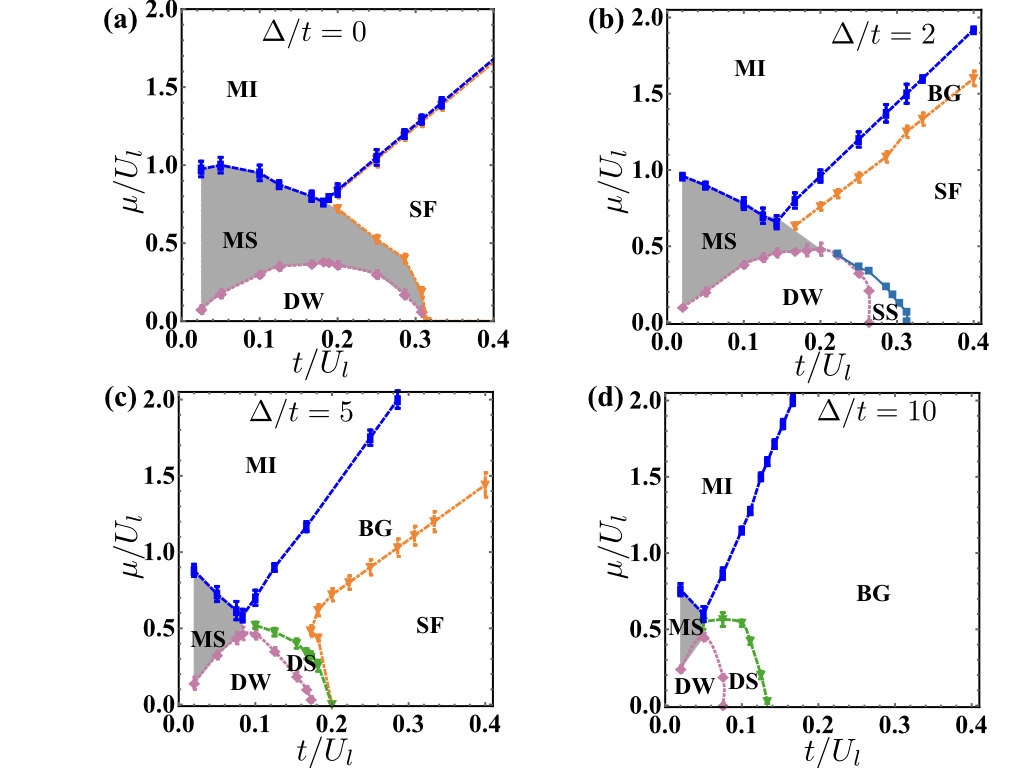}
\caption{Ground state phase diagrams of model~\ref{Eq1} as a function of $t/U_{l}$ and $\mu/U_l$ for clean system (a), and disordered system with disorder strength $\Delta/t=2$ (b), 5 (c), and 10 (d), respectively. The gray shadowed region represents the metastable states. }
\label{FIG4}
\end{figure*}

Figure~\ref{FIG4} shows the ground state phase diagram of hard-core bosons trapped in a two-dimensional optical lattice with cavity-mediated long-range interactions for both clean (Fig.~\ref{FIG4} (a)) and disordered systems (Fig.~\ref{FIG4} (b)-(d)), respectively. The x-axis is $t/U_{l}$ and the y-axis is $\mu/U_{l}$ where $t$ is the hopping amplitude, $U_l$ is the strength of cavity-mediated long-range interactions, and $\mu$ is the chemical potential. Here we set the hopping amplitude $t=1$. The phase boundary is determined by considering cuts through the x axis ($t/U_l$) and calculating the above three order parameters as a function of $\mu/U_l$. Finite-size scaling method is also used to get accurate transition points on the phase boundary.

\begin{figure}[h]
\includegraphics[trim=2.cm 2cm 0cm 2cm, clip=true, width=0.5\textwidth]{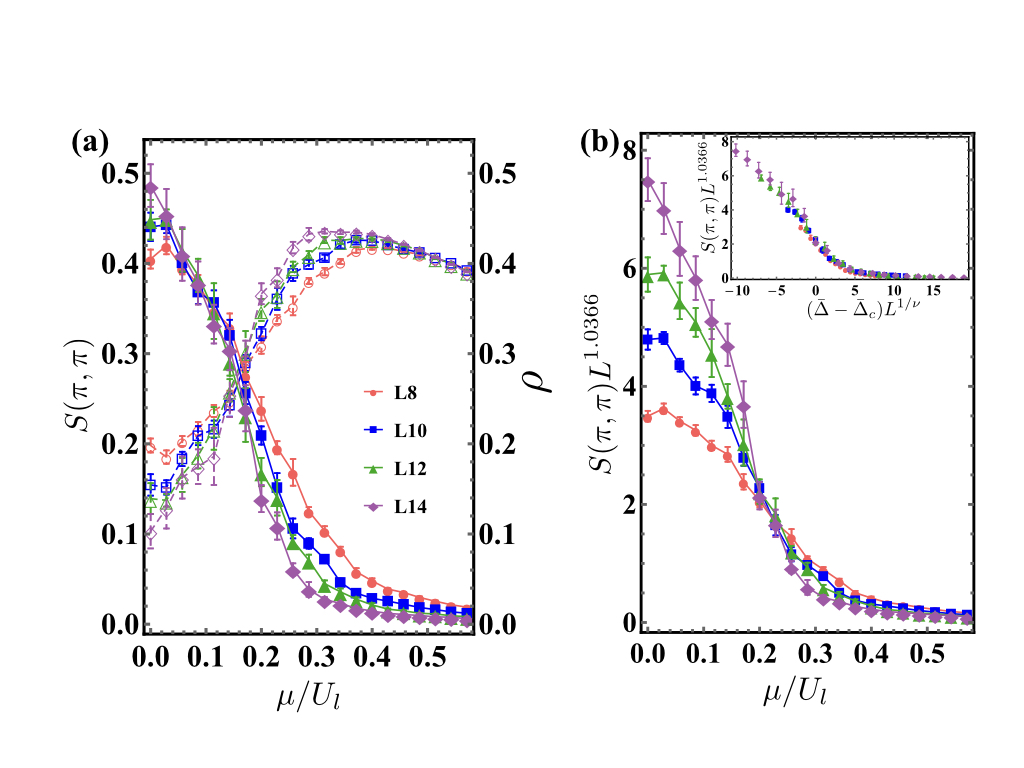}
\caption{Model 1 (cavity-mediated long-range interactions): all following results are at fixed disorder strength $\Delta/t=2$ and $t/U_{l}=0.2857$. (a) The structure factor $S(\pi,\pi)$ and superfluid stiffness $\rho$ as a function of $\mu/U_l$ for different system sizes $L=8$, 10, 12, and 14 (red dots, blue rectangles, green up triangles, and purple diamonds). (b) The finite size scaling result of structure factor $S(\pi, \pi)$ for above system sizes. The crossing of different curves marks a transition point at $\mu/U_{l}=0.2 \pm 0.05$. Insert shows the data collapse result using $\nu=0.67$ and $\bar{\Delta}_c=(t/U_{l})_c=0.2$ corresponding to the critical point extracted from main plot. }
\label{FIG5}
\end{figure}

\begin{figure}[h]
\includegraphics[trim=4.2cm 0cm 0cm 2cm, clip=true, width=0.6\textwidth]{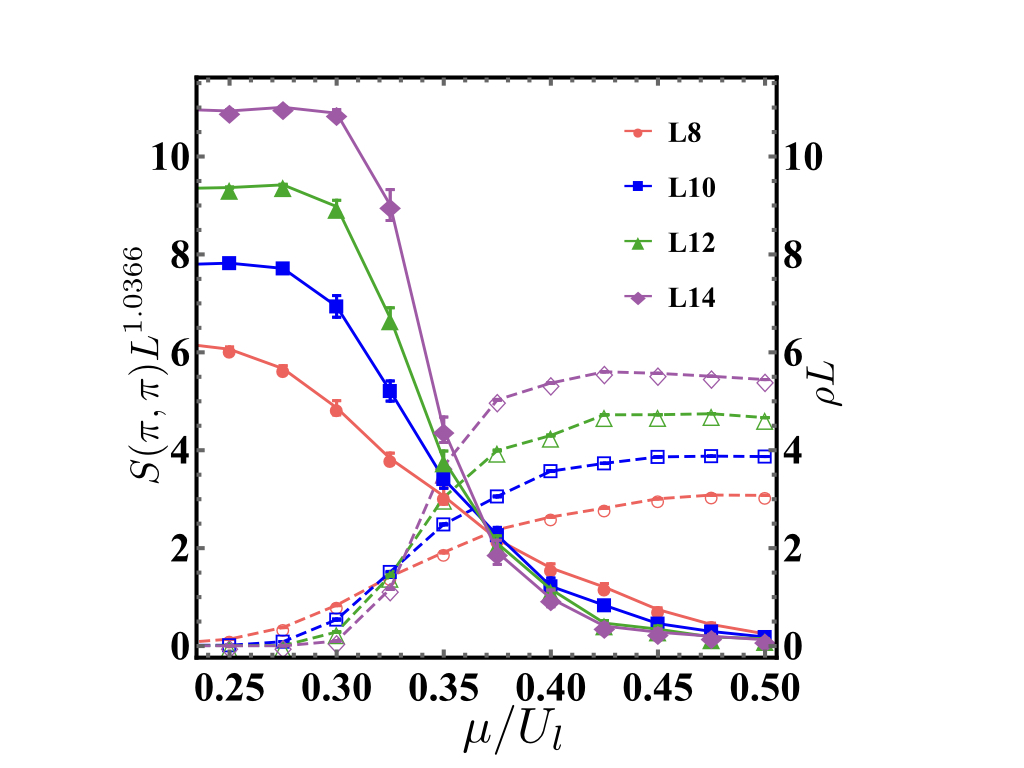}
\caption{Model 1 (cavity-mediated long-range interactions): the scaling of superfluid stiffness $\rho L$ and structure factor $S(\pi, \pi) L^{1.0366}$ as a function of $\mu/U_l$ for different system sizes $L=8$, 10, 12, and 14 (red dots, blue rectangles, green up triangles, and purple diamonds) at $\Delta/t=2$ and $t/U_l=0.25$. Solid lines represent structure factor and dashed lines represent superfluid stiffness. The crossing of different solid curves marks the transition point at $\mu/U_l=0.325 \pm 0.01$ for the density-wave to supersolid phase transition while the crossing of different dashed curves marks the transition point at $\mu/U_l=0.368 \pm 0.02$ for the supersolid to superfluid phase transition.}

\label{FIG7}
\end{figure}

Figure~\ref{FIG4} (a) shows the ground state phase diagram of the clean system. %of the extended BHM with cavity-mediated long-range interaction. 
There are three phases in the phase diagram: a SF phase, a MI phase, and a DW phase. At small $t/U_l$, there is a large region of metastable states (MS). A metastable state is a state that the system resides in but differs from its state of least energy for an extended period of time. %The existence of a large region of metastable state can be understood in the following: the presence of long-range interactions makes decay processes like the phase separation energetically costly, leads to the result that the system stays in the high-energy states for an extended time. 
Here, cavity-mediated long-range interactions tend to induce metastability. Experimentally, the MS can be probed by preparing the system in a well-defined state, providing extra energy, and observing which state the system relaxes to~\cite{Hruby:2018ec}. In the simulation, the MS is determined by starting simulations with two well-defined initial states (DW and MI) and measuring order parameters to determine the final state. If the final state depends on the initial state, the system is in the MS. %If the system with two different initials state can be evolved into two different final states, then the system is in MS. 
Figure~\ref{FIG4} (a) shows that the MS surrounds the DW phase for clean system at a small value of $t/U_l$. The MS and DW persists until $t/U_l=0.32 \pm 0.02$. When $t/U_l>0.32$, the system stays in the SF phase at a lower filling and MI at an integer filling $n=1$. Here, the MI to SF phase transition is the second-order phase transition. %The phase diagram in 1D has been found in~\cite{Igloi:2018ig, Blass:2018iw}. 

Figure~\ref{FIG4} (b) shows the ground state phase diagram at disorder strength $\Delta/t=2$. As disorder is added to the system, the region of MS shrinks. Disorder tends to localize bosons and suppress metastability. Besides the SF, MI and DW phases, there are two new phases emerge: a BG phase and a supersolid (SS) phase. The BG phase intervenes as a Griffiths phase between the MI and SF phases, and it is explained by the theory of inclusions~\cite{Pollet2009BG, Gurarie:2009it}. The SS phase has both diagonal long-range order and off-diagonal long-range order and is characterized by a finite superfluid stiffness $\rho$ and a finite structure $S(\pi, \pi)$. Figure~\ref{FIG5} (a) shows the structure factor $S(\pi, \pi)$ and superfluid stiffness $\rho$ as a function of $\mu/U_l$ for different system sizes $L=8$, 10, 12, and 14 (red dots, blue rectangles, green up triangles, and purple diamonds) at disorder strength $\Delta/t=2$ and $t/U_{l}=0.2857$. Between $0< \mu/U_l < 0.25$, the system is in the SS phase with both a finite superfluid stiffness and a finite structure factor. Figure~\ref{FIG5} (b) shows the finite-size scaling of structure factor $S(\pi, \pi)$, where we plot $S(\pi,\pi)L^{2\beta/\nu}$ as a function of $\mu/U_l$ for system sizes $L=8$, 10, 12, and 14 (the critical exponents $2\beta/\nu$ = 1.0366(8) correspond to the three-dimensional Ising universality class~\cite{Hasenbusch:1999eh}). The crossing of different curves marks the transition point at $\mu/U_{l}=0.2 \pm 0.05$. The insert shows the data collapse result using $\nu=0.67$ and $\bar{\Delta}_c=(\mu/U_{l})_c=0.2$ corresponding to the critical point extracted from main plot. Here, the SS goes to the SF phase via a second-order phase transition which belongs to the (2+1)-dimensional Ising type transition.

Figure~\ref{FIG7} shows the finite-size scaling of superfluid stiffness $\rho$ and structure factor $S(\pi, \pi)$ as a function of $\mu/U_l$ for different system sizes at $\Delta/t=2$ and $t/U_l=0.25$. %The solid lines represent the structure factor and the dashed line represent the superfluid stiffness. 
Dashed lines are the finite-size scaling result of superfluid stiffness. We plot $\rho L^{d+z-2}$ as a function of $\mu/U_{l}$ at $t/U_{l}=0.25$ for a variety of system sizes. Here, $z=1$ is the dynamic critical exponent and the inverse temperature $\beta=L$ is used. The crossing of different curves marks the transition point at $\mu/U_{l}=0.325\pm 0.01$. This shows at $t/U_{l}=0.25$, the DW to SS phase transition is the second-order phase transition and happens at $\mu/U_{l}=0.325\pm 0.01$. Solid lines are the finite-size scaling result of structure factor. The crossing shows that the SS to SF phase transition happens at $\mu/U_l=0.368 \pm 0.02$. At $t/U_l=0.25$, the system is in the SS phase at $0.325 < \mu/U_l < 0.368$.

Figure~\ref{FIG4} (c) shows the ground state phase diagram at disorder strength $\Delta/t=5$. As disorder increases, both the MS and SF phases shrink. This is due to the fact that disorder tends to localize bosons and destroy superfluidity and metastability. Interestingly, we do not find the SS phase but the DS phase. The DS is characterized by a finite compressibility and a finite structure factor but no superfluid stiffness. 

Figure~\ref{FIG4} (d) shows the ground state phase diagram at disorder strength $\Delta/t=10$. At such a strong disorder, there is no SF phase anymore. By comparing phase diagrams in Figure~\ref{FIG1}, we can see that disorder tends to shrink the region of MS states and destroy superfluidity. The BG phase intervenes as a Griffiths phase between the MI and SF phases. The SS or DS intervenes between the DW and SF phases depending on the strength of disorder.

\subsection{Nearest-neighbor interaction}
\label{sec4.2}

\begin{figure*}[th]
\includegraphics[trim=0cm 0cm 0cm 0cm, clip=true, width=0.85\textwidth]{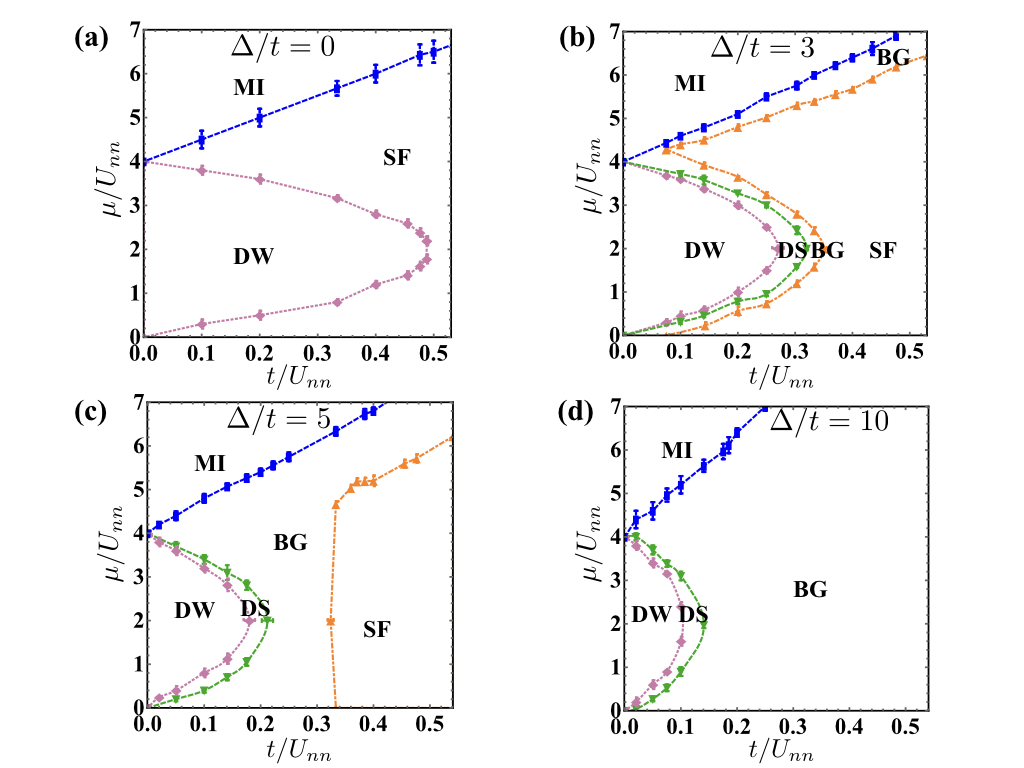}
\caption{Ground state phase diagrams of model~\ref{Eq2} as a function of $t/U_{nn}$ and $\mu/U_{nn}$ for clean system (a), and disordered system with disorder strength $\Delta/t=3$ (b), 5 (c), and 10 (d), respectively.} 
\label{FIG1}
\end{figure*}

In this subsection, we study the ground state phase diagram of hard-core bosons trapped in an optical lattice with nearest-neighbor interactions for both clean (Fig.~\ref{FIG1} (a)) and disordered systems (Fig~\ref{FIG1} (b)-(d)), respectively. The x-axis is $t/U_{nn}$ and the y-axis is $\mu/U_{nn}$, where $t$ is the hopping amplitude, $U_{nn}$ is the strength of nearest-neighbor interactions, and $\mu$ is the chemical potential. The hopping amplitude is set to $t=1$. The phase boundary is determined by using system size $L=16$, finite size scaling method is also used to get accurate transition points on the phase boundary.%There are various phases exist in the above phase diagrams. Table~\ref{Table1} shows the quantum phases in Figure~\ref{FIG1} and the corresponding order parameters: superfluid stiffness $\rho$, structure factor $S(\pi, \pi)$, and compressibility $\kappa$. 

Figure~\ref{FIG1} (a) shows the ground state phase diagram of the extended BHM with nearest-neighbor interactions for the clean system. There are three phases in the phase diagram: a SF phase, a MI phase, and a DW phase. Here, the MI to SF phase transition is the second-order phase transition at an integer filling while the DW to SF phase transition is the first-order phase transition at a half filling~\cite{Batrouni:2000cd}. 

\begin{figure}[h]
\includegraphics[trim=2.cm 2cm 0cm 0cm, clip=true, width=0.55\textwidth]{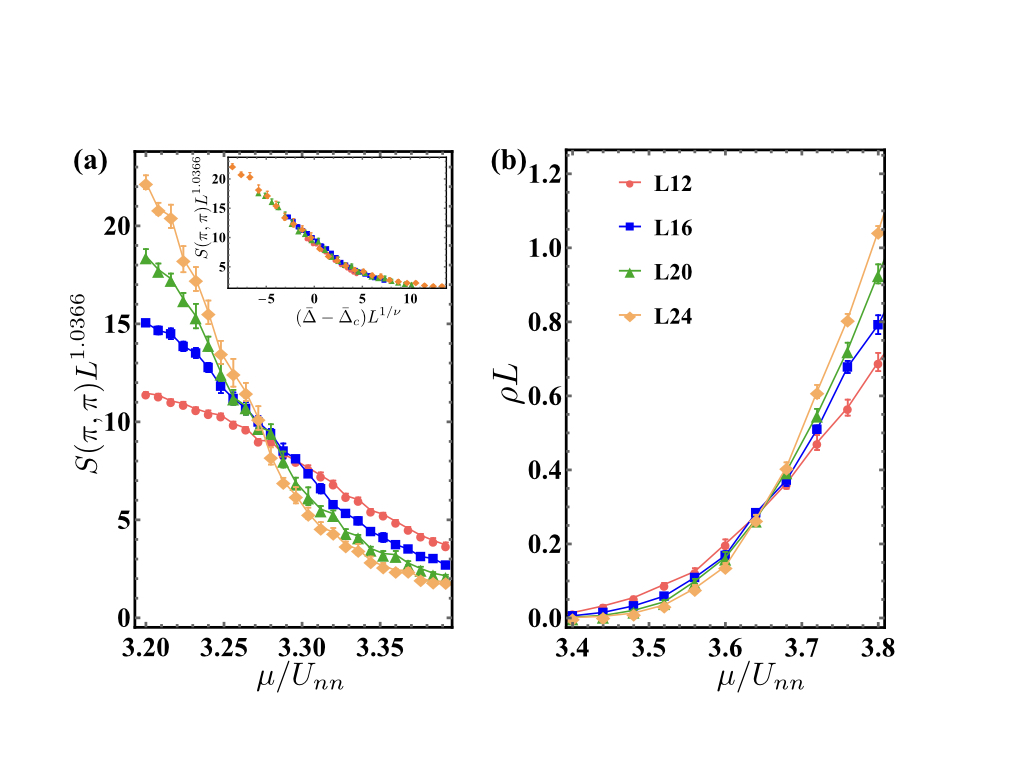}
\caption{Model 2 (nearest-neighbor repulsive interactions): all following results are at fixed disorder strength $\Delta/t=3$ and $t/U_{nn}=0.2$. (a) Main plot: finite size scaling of $S(\pi, \pi)$ for system sizes $L=12$, 16, 20, and 24 (red circles, blue rectangles, green up triangles, and orange diamonds, respectively). The crossing of different curves marks the transition point at $\mu/U_{nn}=3.275 \pm 0.02$ for the DS to BG phase transition. Insert shows the data collapse result using $\nu=0.67$ and $\bar{\Delta}_c=(\mu/U_{nn})_c=3.275$ corresponding to the critical point extracted from main plot. (b) shows the scaling of superfluid stiffness $\rho L$ as a function of $\mu/U_{nn}$ for $t/U_{nn}=0.2$ for $L=12$, 16, 20, and 24. The crossing of different curves marks the transition point at $\mu/U_{nn}=3.652\pm 0.02$ for the BG to SF phase transition.}
\label{FIG2}
\end{figure}

Figure~\ref{FIG1} (b) shows the ground state phase diagram at disorder strength $\Delta/t=3$. As disorder is added to the system, the glassy phase (BG and DS) emerge. This can be explained by the theory of inclusions~\cite{Pollet2009BG, Gurarie:2009it}, which states that a compressible glassy phase surrounds the incompressible phase. There are two kinds of glassy phase, the BG phase and DS phase. The BG phase is characterized by a finite compressibility $\kappa$ but zero structure factor $S(\pi, \pi)$ while the DS phase is characterized by both a finite compressibility and a finite structure factor. The incompressible phase here is the DW phase. As $t/U_{nn}$ increases, the system goes from the DW to DS phase transition and then the DS to BG phase transition. The DS phase intervenes between the DW and BG phase since both have a finite structure factor. Finite-size scaling method is used to determine all critical points on the phase boundary. As shown in Figure~\ref{FIG2}, at fixed $t/U_{nn}=0.2$, as $\mu/U_{nn}$ increases, the DW phase goes to the DS phase first. The main plot of Figure~\ref{FIG2} (a) shows the finite-size scaling of $S(\pi, \pi)$ at fixed disorder strength $\Delta/t=3$ and $t/U_{nn}=0.2$ for system sizes $L=12$, 16, 20, and 24 (red circles, blue rectangles, green up triangles, and orange diamonds, respectively). The crossing of different curves marks the transition point at $\mu/U_{nn}=3.275 \pm 0.02$. Insert shows the data collapse result using $\nu=0.67$ and $\bar{\Delta}_c=(\mu/U_{nn})_c=3.275$ corresponding to the critical point extracted from main plot, which shows the DS to BG phase transition belongs to the (2+1)-dimensional Ising type transition.  Figure~\ref{FIG2} (b) shows the scaling of superfluid stiffness $\rho L^{d+z-2}$ with $z=1$, as a function of $\mu/U_{nn}$ for $t/U_{nn}=0.2$ and $L=12$, 16, 20, and 24. Here, $z$ is the dynamic critical exponent and the inverse temperature $\beta=L$ is used. The crossing of different curves marks the transition point at $\mu/U_{nn}=3.652\pm 0.02$, which shows that at $t/U_{nn}=0.2$, the BG to SF phase transition is the second-order phase transition and happens at $\mu/U_{nn}=3.652\pm 0.02$.

\begin{figure*}[h]
\includegraphics[trim=0.cm 2cm 0cm 5cm, clip=true, width=0.8\textwidth]{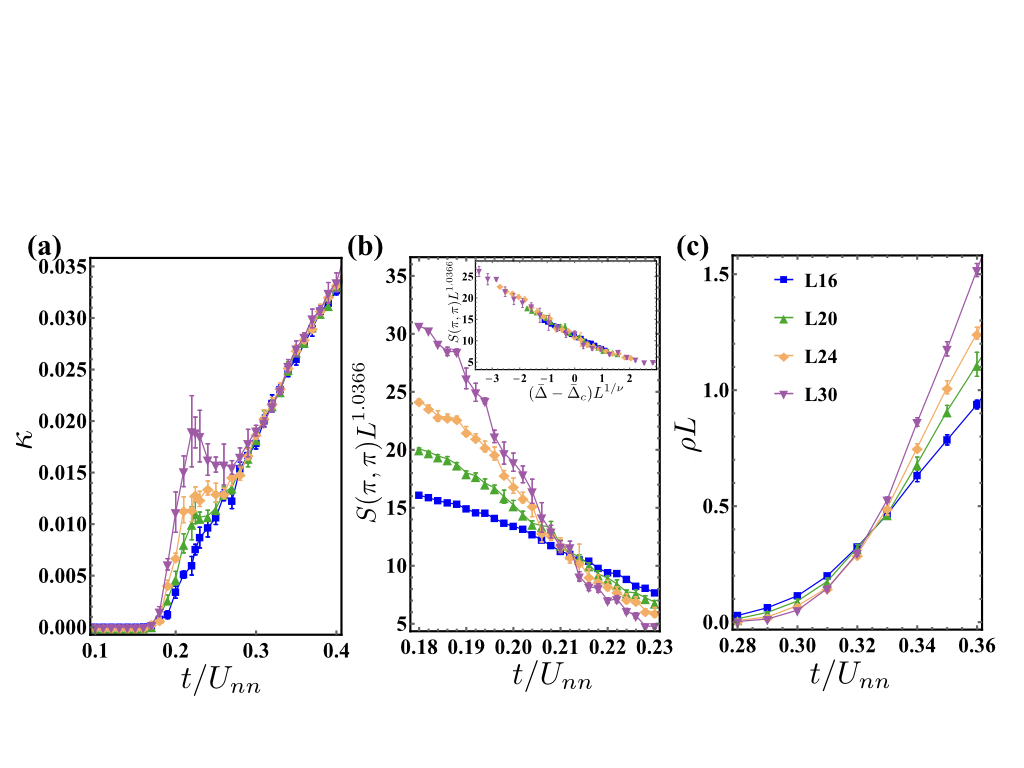}
\caption{Model 2 (nearest-neighbor repulsive interactions): all following results are at fixed disorder strength $\Delta/t=5$ and $\mu/U_{nn}=2$. (a) Compressibility $\kappa$ as a function of $t/U_{nn}$ for different system sizes $L=16$, 20, 24, and 30 (blue rectangles, green up triangles, orange diamonds, and purple down triangles, respectively). The DW to DS phase transition happens around $t/U_{nn}=0.18 \pm 0.01$. (b) Finite size scaling result of $S(\pi, \pi)$ for above system sizes. The crossing of different curves marks the transition point at $t/U_{nn}=0.212 \pm 0.01$ for the DS to BG phase transition. Insert shows the data collapse result using $\nu=0.67$ and $\bar{\Delta}_c=(t/U_{nn})_c=0.212$ corresponding to the critical point extracted from main plot. (c) Finite size scaling of superfluid stiffness $\rho L$ as a function of $t/U_{nn}$  for above system sizes. The crossing of different curves marks the transition at $t/U_{nn}=0.3241\pm 0.005$ for the BG to SF phase transition.}
\label{FIG3}
\end{figure*}

Figure~\ref{FIG1} (c) shows the ground state phase diagram at disorder strength $\Delta/t=5$. As disorder increases, the SF phase shrinks. This is because disorder tends to localize bosons and destroy superfluidity. The MI and DW phase also shrinks and we have a large region of BG phase. Figure~\ref{FIG3} shows the results at disorder strength $\Delta/t=5$ and $\mu/U_{nn}=2$. At a fixed filling factor $n=0.5$, as $t/U_{nn}$ increases, the DW phase is unstable and the system goes to the DS phase. Figure~\ref{FIG3} (a) shows the compressibility $\kappa$ as a function of $t/U_{nn}$ for different system sizes $L=16$, 20, 24, and 30 (blue rectangles, green up triangles, orange diamonds, and purple down triangles, respectively). The compressibility $\kappa$ stays zero until $t/U_{nn}=0.18$ and then becomes finite, which shows the DW to DS phase transition happens around $t/U_{nn}=0.18 \pm 0.01$. As $t/U_{nn}$ increases further, the DS is unstable and the system enters the BG phase. Figure~\ref{FIG3} (b) shows the finite-size scaling of $S(\pi, \pi)$  for above system sizes. The crossing of different curves marks the transition point at $t/U_{nn}=0.212 \pm 0.01$. Insert shows the data collapse result using $\nu=0.67$ and $\bar{\Delta}_c=(t/U_{nn})_c=0.212$ corresponding to the critical point extracted from main plot. This shows the DS goes to the BG phase via a second-order phase transition which belongs to the (2+1)-dimensional Ising type transition. Finally, the BG goes to the SF as $t/U_{nn}$ increases further. Figure~\ref{FIG3} (c) shows the finite-size scaling of superfluid stiffness $\rho L$ as a function of $t/U_{nn}$ for above system sizes. The crossing of different curves marks the transition point at $t/U_{nn}=0.3241\pm 0.005$.

Figure~\ref{FIG1} (d) show the ground state phase diagram at disorder strength $\Delta/t=10$. At such a strong disorder, there is no SF phase anymore. By comparing phase diagrams in Fig.~\ref{FIG1}, we can see that disorder tends to localize bosons and destroy superfluidity. Compared to the disordered BHM without nearest-neighbor interactions~\cite{Soyler:2011ik}, we find that the DW phase can not go to the BG phase directly, there is the DS phase intervenes between them.

\section{Conclusion}
\label{sec:sec5}

In this paper, we use quantum Monte Carlo simulations with the worm algorithm to study the phase diagram of a two-dimensional Bose-Hubbard model with cavity-mediated long-range interactions for both clean and disordered systems in the hard-core limit. We find the SS phase at a weak disorder and DS phase at a stronger disorder. Due to the long-range interaction term, we find a large region of metastable states in both clean and disordered systems. Disorder suppresses metastable states and superfluidity. We also study the phase diagram of the extended Bose-Hubbard model with nearest-neighbor interactions for both clean and disordered systems in the hard-core limit. We find that there are two kinds of glassy phases: the BG phase and the DS phase. The glassy phase intervenes as a Griffiths phase between the DW and SF phases, which can be explained by the theory of inclusions. The DS phase intervenes between the DW and BG phase since both the DS and DW phases have a finite structure factor.

{\textit{Acknowledgements}}  
This work was performed with financial support from Saarland University. We would like to thank B. Capogrosso-Sansone for enlightening discussions. The computing for this project was performed at the OU Supercomputing Center for Education $\&$ Research (OSCER) at the University of Oklahoma (OU) and the cluster at Saarland University.

\bibliography{hardcore}

%\begin{thebibliography}{50}
%\bibitem{Santos} X. Deng, R. Citro, E. Orignac, A. Minguzzi, and L. Santos\njp{15}{045023}{2013}.
%\end{thebibliography}

\end{document}